# Enhanced absorption in heterostructures with graphene and truncated photonic crystals

Yiping Liu[1], Lei Du[1], Yunyun Dai[2], Yuyu Xia[2], Guiqiang Du[1,*] Guang Lu[1], Fen Liu[1], Lei Shi[2], Jian Zi[2]

[1] *School of Space Science and Physics, Shandong University at Weihai, Weihai 264209, China*

[2] *Department of Physics, Key Laboratory of Micro- and Nano-Photonic Structures (Ministry of Education) and State Key Laboratory of Surface Physics, Fudan University, Shanghai 200433, China*

Abstract

We theoretically and experimentally investigate the optical absorption properties of heterostructures composed of graphene films and truncated photonic crystals (PCs) in the visible range. The experimental results show that the absorption of the heterostructure is enhanced greatly in the whole forbidden gap of PCs compared with that of graphene alone. The absorption is enhanced over a wide angle of incidence for both transverse electric (TE) and transverse magnetic (TM) polarizations. The enhanced absorption band broadens for TE polarization but narrows for TM polarization as the incident angle increases. In the forbidden gap of the PCs, the maximum absorptance of the heterostructures is nearly four times of that of bare graphene films for arbitrary incident angles and polarizations. The optical experiments are in excellent agreement with the theoretical results.

Graphene has attracted huge interest as an ultra-thin gapless two-dimensional semiconductor[1] because of its unique electrical and optical properties. These

---

*Corresponding author: *dgqql@sdu.edu.cn*

properties have led to numerous important applications in optoelectronic and nanophotonic devices such as transparent electrodes,[2] ultrafast lasers,[3] polarizers,[4] and photodetectors.[5] A unique optical property of graphene is that its absorption is nearly constant over the visible and near-infrared regions.[6-8] For example, the absorptance of a graphene monolayer is about 2.3% at the normal incidence of light, and is almost independent of wavelength.[7,8]

Although the absorption of graphene remains nearly in variable over a broad wavelength range, its absorptance is still small, which makes graphene a good candidate to fabricate optoelectronic devices and components such as transparent electrodes and optical display materials.[2,9,10] However, the low absorptance of graphene limits its applications in optoelectronic devices that need strong light–graphene interactions, such as graphene-based photodetectors[5,11,12] and optical absorbers. Therefore, many methods have been proposed to enhance the absorption of graphene including exciting plasmons in doped graphene nanostructres,[13] combining graphene with conventional plasmonic nanostructures,[14,15] forming periodically patterned graphene,[16] and integrating graphene with Fabry–Pérot microcavities.[11,17,18] Most of these approaches realized enhanced absorption of graphene at specific wavelengths or narrow region. However, it is necessary to achieve absorption enhancement of graphene over a wide wavelength range.

Compared with the above methods to enhance the absorption of graphene using nanostructures or by putting graphene in microcavities under specific technological conditions, it is relatively easy to prepare graphene films on the top of bulk materials or one-dimensional photonic crystal (PCs). An important theoretical work proposed that enhanced absorption of graphene over the whole forbidden gap of the PC is realized in sandwich structures composed of a graphene film, spacer layer and truncated PC.[19] In fact, mono- and few-layer graphene films are highly transparent in the visible region, so they don't behave as good optical barriers. In this paper, we deposit monolayer, bilayer and trilayer graphene on the top of truncated dielectric PCs. Our theoretical and experimental results show that enhanced absorption of graphene is realized over the whole forbidden gap of the PC.

The heterostructures considered here are composed of a graphene film and one-dimensional truncated PC made of $SiO_2$ and $TiO_2$. The heterostructure is denoted as $G(AB)^N AS$, as shown in Fig. 1, where G represents the graphene film, A denotes $SiO_2$ with refractive index $n_A=1.431$[20] and thickness $d_A=98.4$ nm, B is $TiO_2$ with $n_B=2.123$[20] and $d_B=66.4$ nm, and N is the periodic number. The substrate S is BK7 with $n_S=1.51$. The refractive index of graphene is described by a simple model $n_G=3.0+C_1\lambda/3i$ ($C_1=5.446$ μm$^{-1}$) in the visible range,[21] and the thickness of a graphene monolayer is 0.34 nm. The transfer matrix method[22] was used to calculate the transmittance ($T$) and reflectance ($R$) of the heterostructures, and their absorptance ($A$) was calculated using the formula $A=1-T-R$. The alternating A and B layers of truncated $SiO_2/TiO_2$ PCs with the structure $(AB)^N A$ were deposited on the planar face of the substrate (S) by an ion-assisted electron-beam evaporation under high vacuum. The thicknesses of A and B layers were monitored by a quartz crystal sensor. A monolayer graphene grown by chemical vapor deposition on Cu foils was placed on the surface of $(AB)^N AS$ by a wet transfer method.[15] The wet transfer process is repeated to realize bilayer and trilayer graphene on the top of truncated PCs. *T* and *R* of the heterostructure samples were measured by the Cary-100 ultraviolet–visible–near-infrared spectrophotometer and R1-angle-resolved spectrometer system.

Firstly, the normal incidence of the light is considered. Figure 2 (a)-(c) display absorption spectra of the heterostructure $G(AB)^{19}AS$ with graphene films of different thicknesses, where $G_1$, $G_2$ and $G_3$ denote monolayer, bilayer and trilayer graphene, respectively. The red and blue lines describe the theoretical and experimental results, respectively. The spectra show that the absorption of the graphene films is enhanced strongly in the whole forbidden gap of the PC. The measured maximum absorptance of the heterostructure with monolayer, bilayer and trilayer graphene is 0.0866, 0.1692, and 0.2447 at 568.3 nm, respectively, and the corresponding simulated values at 568.3 nm are 0.0889, 0.1699, and 0.2439, respectively. The experimental spectra are in good agreement with the theoretical ones. Compared with that of bare monolayer, bilayer or trilayer graphene, the absorptance of the heterostructure containing the graphene layers is increased by nearly four times.

To reveal the physical origin of the enhanced absorption in the heterostructures containing graphene, we numerically calculated the electric field intensity distribution $|E|^2$ for the heterostructure $G(AB)^{19}AS$ with a graphene monolayer at $\lambda=568.3$nm. The results are presented in Fig. 3. The thickness of the graphene monolayer in the schematic diagram is 100 times its actual thickness. The electric field intensity of the incident wave is supposed to be 1. The electric field intensity in the graphene monolayer is nearly four times of that of the incident wave. This is because the PC behaves as a reflecting mirror in its forbidden gap, which results in strong photon localization on the surface of the truncated PC. Therefore, enhanced electric field intensity appears in the graphene monolayer, which causes the enhanced absorption. In contrast, the electric field intensity in the bare graphene monolayer is nearly 1.

Moreover, the oblique incidence of the light was considered for both transverse electric (TE) and transverse magnetic (TM) polarizations. Figure 4 and 5 display the absorption spectra as a function of wavelength for TE and TM waves at oblique incidence, respectively where (a)-(c) show 30° incident angle, and (d)-(f) depict 60° incident angle.

Figure 4 (a)-(c) present the theoretical (red lines) and experimental (blue lines) absorption spectra of the heterostructure G(AB)[19]AS for a TE wave at 30° incident angle. The enhanced absorption band of the heterostructure is broadened because the forbidden gap of PC is widened as the incident angle for TE polarization increases. The measured maximum absorptance of the heterostructures with monolayer, bilayer and trilayer graphene is 0.1028, 0.1939, and 0.2733 at 540.2 nm, respectively. These values are larger than those at normal incidence. The theoretical absorptance of the heterostructures with monolayer, bilayer and trilayer graphene at 540.2 nm is 0.1019, 0.1936, and 0.2760, respectively. The experimental spectra are consistent with the theoretical ones. Compared with that of bare graphene at the same incident angle, the absorptance of the heterostructure containing the monolayer, bilayer or trilayer graphene is still enhanced by nearly four times. The physical origin of the enhanced absorption in the heterostructure at an incident angle of 30° is similar to that at normal incidence.

Figure 4 (d)-(f) show the enhancement of broader absorption band of the heterostructure G(AB)$^{19}$AS for a TE wave at 60° incident angle is larger than that at 30° incident angle. This is because the PC has a wider forbidden gap at larger incident angle. The experimental maximum absorptance of the heterostructures with monolayer, bilayer and trilayer graphene is 0.1716, 0.3049, and 0.4141 at 481.8 nm, respectively. These values are larger than the corresponding ones at 30° incident angle. The theoretical absorptance of the heterostructures with monolayer, bilayer and trilayer graphene at 481.8 nm is 0.1700, 0.3115, and 0.4299, respectively. Compared with that of a bare graphene film at the same incident angle, the absorptance of the heterostructures containing graphene is enhanced by nearly four times. The experimental spectra agree well with the numerical values.

Theoretical (red lines) and experimental (blue lines) absorption spectra of the heterostructure G(AB)$^{19}$AS for a TM wave at 30° incident angle are presented in Fig. 5 (a)-(c). Different from those for TE polarization, the enhancement of the absorption band of the heterostructure is smaller than that at normal incidence because the forbidden gap of the PC is narrowed with increasing incident angle for TM polarization. The experimental maximum absorptance of the heterostructures with monolayer, bilayer and trilayer graphene is 0.0772, 0.1488, and 0.2158 at 548.5 nm, respectively. These values are smaller than those at normal incidence. The calculated absorptance of the heterostructures with monolayer, bilayer and trilayer graphene at 548.5 nm is 0.0774, 0.1489, and 0.2150, respectively. Compared with a bare graphene film at the same incident angle, the absorptance of the heterostructures containing graphene is enhanced by nearly four times. Again, the experimental spectra are in good agreement with the simulated results.

Figure 5 (d)-(f) show that the width of enhanced absorption band of the heterostructure $G(AB)^{19}AS$ for a TM wave at 60° incident angle is shortened compared with that at 30° incident angle. This is because the forbidden gap of the PC is narrower at larger incident angle. The experimental maximum absorptance of the heterostructures with monolayer, bilayer and trilayer graphene is 0.0482, 0.0849, and 0.1242 at 484.6 nm, respectively. These absorptance values are smaller than the corresponding ones at 30° incident angle. The theoretical absorptance of the heterostructures with monolayer, bilayer and trilayer graphene at 484.6 nm is 0.0455, 0.0889, and 0.1303, respectively. Compared with that of a bare graphene film at the same incident angle, the absorptance of the heterostructures containing graphene is again enhanced by nearly four times.

The above theoretical and experimental results demonstrate the width of the enhanced absorption band of the graphene film on the top of the PC is determined by that of the forbidden gap of the PC. The enhanced absorption band is broadened with increasing incident angle for TE polarization. This indicates that if an omnidirectional and ultra-wide photonic forbidden gap can be introduced by using PCs composed of high-refractive-index materials,[23] PC heterostructures,[24] or metamaterials,[25] an omnidirectional or ultra-wide enhanced absorption band of graphene will be achieved. These properties are important to fabricate graphene-based broadbandoptoelectronic devices.

In conclusion, we studied the optical absorption of monolayer, bilayer and trilayer graphene deposited on the top of truncated PCs. The width of the enhanced absorption band of graphene is determined by the width of the whole forbidden gap of the PC. Normal and oblique incidences for both TE and TM polarizations were also investigated. The maximum absorptance of the graphene film in the forbidden gap of the PC is nearly four times of that of the equivalent bare graphene film. The experimental results agree well with theoretical values.

This work is supported by the National Natural Science Foundation of China (No. 11264003), the Natural Science Foundation of Shandong Province (No. ZR2015AM008), and the Basic Research Program of Shandong University at Weihai (No. 2015ZQXM013).

Figure captions

FIG. 1. Schematic diagram of a heterostructure G(AB)$^N$AS composed of a graphene film G and photonic crystal (AB)$^N$AS, where A is SiO$_2$, B is TiO$_2$, S is the substrate, and N is periodic number.

FIG. 2. Absorption spectra of the heterostructure G(AB)$^{19}$AS with (a) monolayer graphene G$_1$, (b) bilayer graphene G$_2$, and (c) trilayer graphene G$_3$. A denotes SiO$_2$ with $n_A$=1.431 and $d_A$=98.4 nm, and B is TiO$_2$ with $n_B$ =2.123 and $d_B$=66.4 nm. S is BK7 with $n_S$=1.51. The red and blue lines represent the numerical and experimental results, respectively.

FIG. 3. Simulated electric field intensity |E|$^2$ in the heterostructure G(AB)$^{19}$AS at 568.3 nm and normal incidence. The thickness of the graphene monolayer in the simulation was 100 times its real thickness. Other parameters are the same as those in Fig. 2(a).

FIG. 4. Simulated (red lines) and measured (blue lines) absorption spectra of G(AB)$^{19}$AS as a function of wavelength for a TE wave at incident angles of (a)-(c) 30°and (d)-(f) 60°. Other parameters are the same as those in Fig. 2.

FIG. 5. Simulated (red lines) and experimental (blue lines) absorption spectra of G(AB)$^{19}$AS as a function of wavelength for a TM wave at incident angles of (a)-(c) 30° and (d)-(f) 60°. Other parameters are the same as those in Fig. 2.

Figures

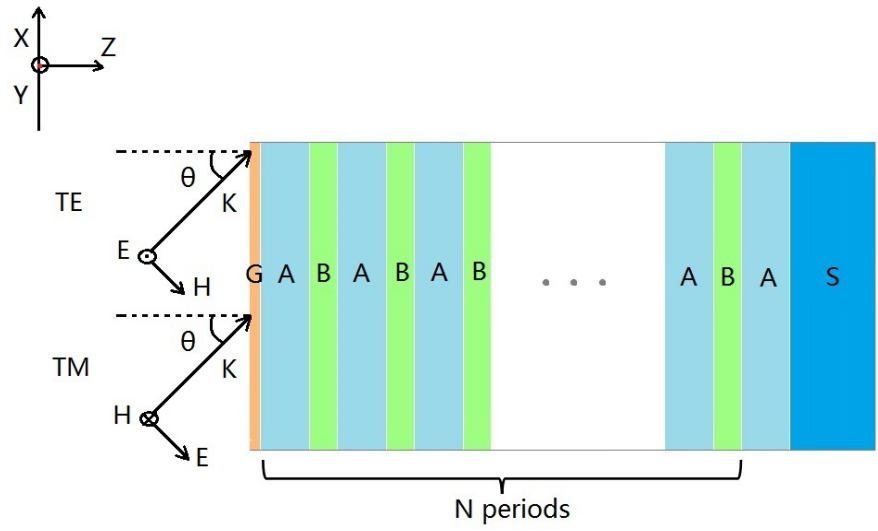

FIG. 1

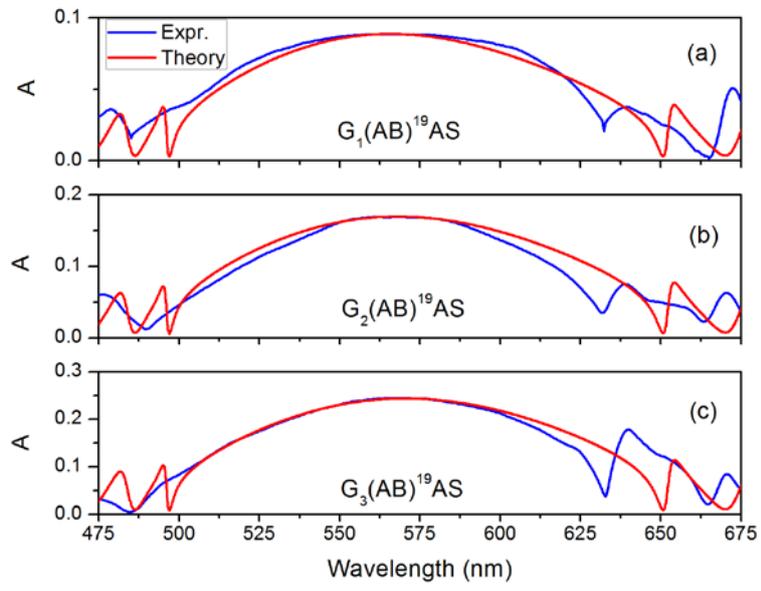

FIG. 2

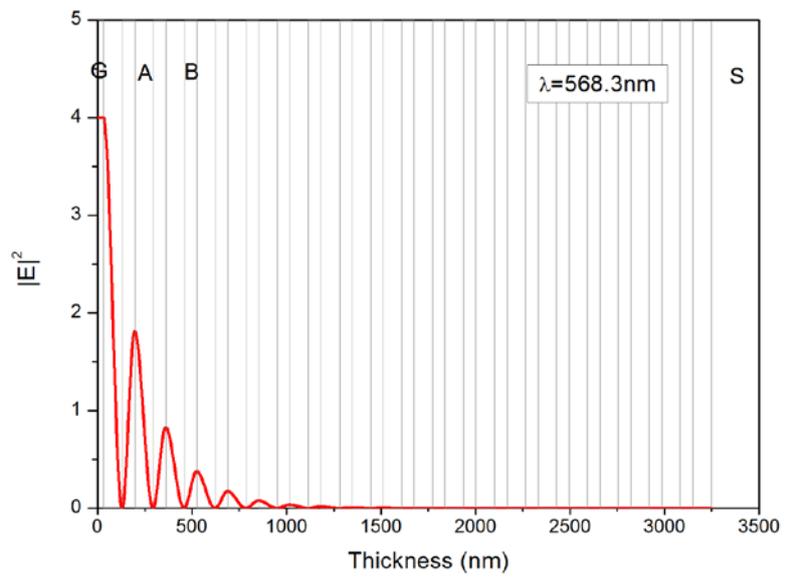

FIG. 3

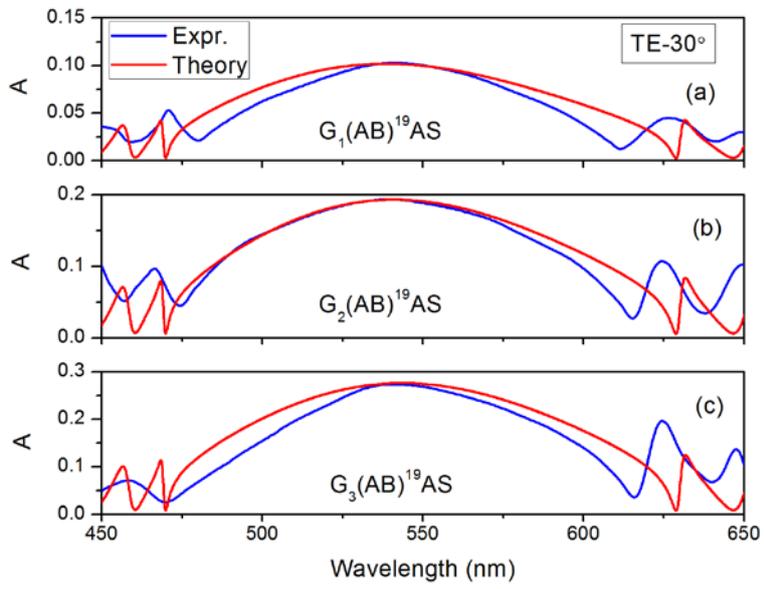

FIG. 4 (a)

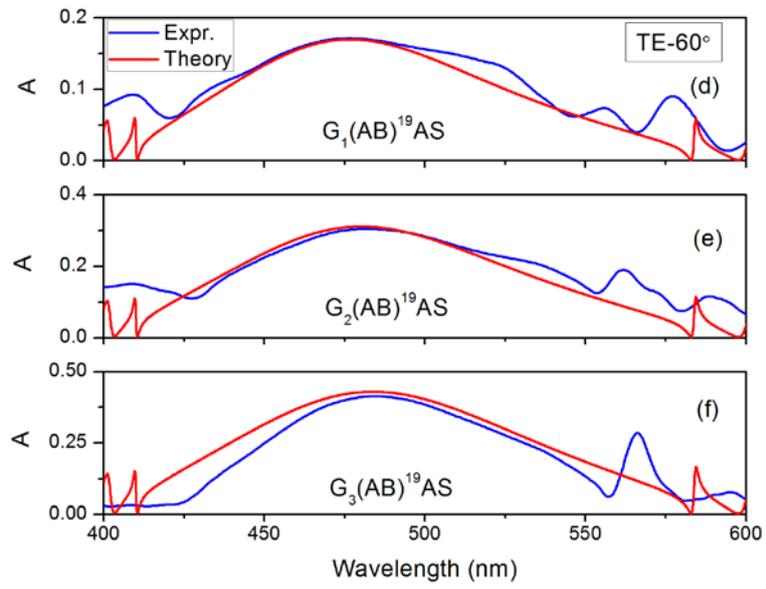

FIG. 4 (b)

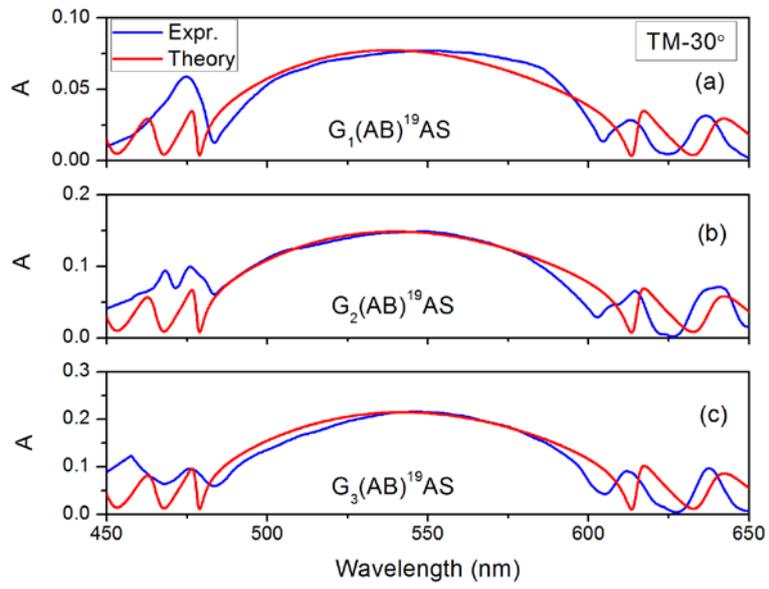

FIG. 5 (a)

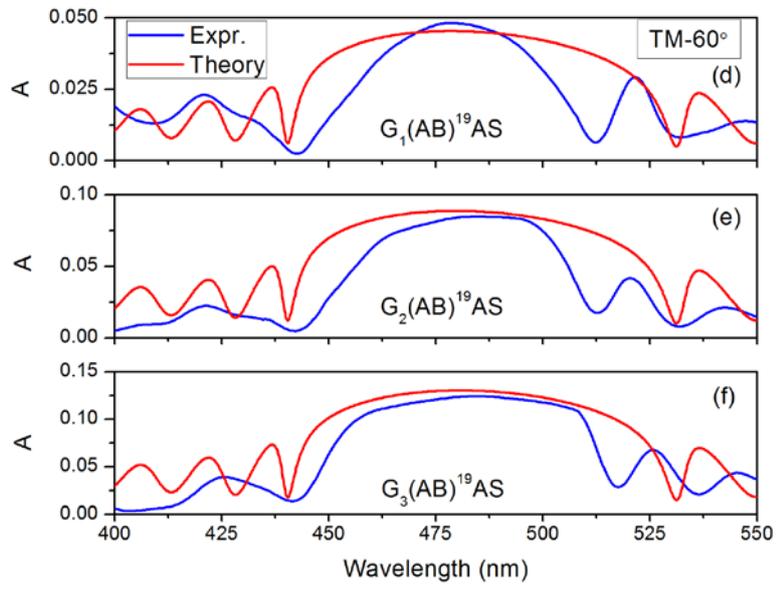

FIG. 5 (b)